# Optical Propulsion and Levitation of Metajets

Kaushik Kudtarkar[1], Yixin Chen[2], Ziqiang Cai[3], Preston Cunha[1], Xinyi Wang[2], Sam Lin[2], Zi Jing Wong[2,4], Yongmin Liu[3,5], and Shoufeng Lan[1,2,6*]

[1] Department of Mechanical Engineering, Texas A&M University, College Station, TX, USA.
[2] Department of Material Science and Engineering, Texas A&M University, College Station, TX, USA.
[3] Department of Electrical & Computer Engineering, Northeastern University, Boston, MA, USA.
[4] Department of Aerospace Engineering, Texas A&M University, College Station, TX, USA.
[5] Department of Mechanical Engineering & Industrial Engineering, Northeastern University, Boston, MA, USA.
[6] Department of Electrical & Computer Engineering, Texas A&M University, College Station, TX, USA.

Corresponding Author: shoufeng@tamu.edu

## Abstract

The quintessential hallmark distinguishing metasurfaces from traditional optical components is the engineering of subwavelength meta-atoms to manipulate light at will. Enabling this freedom, in a reverse manner, to control objects constituted by metasurfaces could expand our capability of optical manipulation to go beyond the predominant microscopic and sub-microscopic scales. Here, we introduce a driving metaphotonic force fully controllable by meta-atoms to manipulate structured objects named metajets. Upon Newton's law of motion that can apply to classical and relativistic mechanics, we develop a first-principles theory to analyze optical forces generated by refraction and reflection at an interface. We find that three-dimensional motions of metajets would be possible if one could introduce an extra wavevector component. We achieve that by creating a spatially distributed phase gradient with deliberately arranged silicon nanopillars. Our experiments and simulations reveal an in-plane propulsion and, very importantly, out-of-plane levitation of the metajets, aligning well with the theory. We also find that the metaphotonic force augments with increased light power but is not limited by the size of metajets, which could unleash new opportunities for metaphotonic control in large settings, such as interstellar light sails.

## Introduction

The motion of objects across the atomic to macroscopic scales, such as electrons, particles, and vehicles, is at the heart of modern transport phenomena. Underline the motion is a force ($F$) that arises from the change of momentum ($q$) with $F = dq/dt$ or $F = (q_\text{out} - q_\text{in})/\Delta t$ within a time interval of $\Delta t$ governed by Newton's law. This first-principles relation of force and momentum holds for classical and relativistic mechanics, including that for massless photons at the speed of light ($c$). In an idealized circumstance, when a photon in a vacuum impinges at a normal incidence angle on a semi-infinite medium, we can assume the output momentum ($q_\text{out}$) is zero. The force then simplifies to $F = -q_\text{in}/\Delta t$, determined by input momentum ($q_\text{in}$) only. Substituting $q_\text{in}$ with $E/c$, where $E$ is the photon energy, we can get $F = -E/(c\Delta t)$. For a light beam with $M$ photons, because $ME/\Delta t$ is the power ($p$) of light, the force further simplifies to $F = -p/c$. The beauty of these simplifications in the idealized circumstance is that $p$ is a routinely measured property.

A more practical scenario happens at an interface where we can treat it as combining two semi-infinite media. As a result, we can rewrite the force as $F_{total} = [(q_{out} - 0) + (0 - q_{in})]/\Delta t$. While adding zeros does not add any values to the force, it empowers us to separate the force as $F_{total} = F_{out} + F_{in}$, in which the output ($F_{out}$) and input ($F_{in}$) forces possess opposite signs. We can define the sign as positive (negative) for light that shines to (from) the interface because $F_{in}$ ($F_{out}$) is parallel (antiparallel) to the wavevector of $k_{in}$ ($k_{out}$) or light propagation direction. Following this sign convention, we can determine $F_{in} = np/c$ by considering the refractive index of the medium ($n$) since $F_{in}$ strictly follows the same format as that for a semi-infinite medium. Subsequently, we can readily know $F_{total}$ if we provide the relationship between $q_{out}$ and $q_{in}$, recognizing that $F_{out}$ and $F_{in}$ are linearly proportional to the two momenta. As momentum locks with wavevector ($k$), the relationship between $q_{out}$ and $q_{in}$ manifests in the classical Snell's law ($k_{out}^{\parallel} - k_{in}^{\parallel} = 0$)[1]. It is crucial to know that $k_{out}^{\parallel}$ and $k_{in}^{\parallel}$ are in-plane components of wavevectors for the output and input light with precisely the same amplitude. Consequently, regular optical refraction and reflection cannot induce in-plane forces, leading to zero lateral motion of objects, such as solar sails with light reflected from mirror-like thin films[2]. On the other hand, if one could induce extra in-plane wavevector components, lateral and three-dimensional motions would be possible.

We tackle the problem by creating an effective wavevector ($k_\phi$) via spatially distributed phase gradient along the x-axis ($k_\phi = \frac{d\phi}{dx}$), leading to a modified Snell's law ($k_{out}^{\parallel} - k_{in}^{\parallel} = k_\phi$) conventionally used to manipulate anomalous refraction and reflection[3-10]. For simplicity, while not compromising the physics, we use a normal incidence light ($k_{in}^{\parallel} = 0$), and the refractive indices for the input and output light are the same as $n$. The modified Snell's law then reduces to $\sin\theta = \frac{\lambda_0}{2n\pi}\frac{d\phi}{dx}$, where $\lambda_0$ is the wavelength of light, and $\theta$ is the angle of the output light. Following our previous analysis, the in-plane force ($F_x$) is a $\sin\theta$ component of that with output light at a normal angle. Thus, we determine that $F_x = -\frac{\eta p \lambda_0}{2\pi c}\frac{d\phi}{dx}$, where η represents the output efficiency. The negative sign follows our sign convention for output light, indicating that the in-plane force acts in a direction opposite to the increasing phase gradient. Besides this in-plane force that applies to anomalous refraction and reflection, we find an out-of-plane force ($F_z$) that differs between them. We obtain that $F_z = \frac{np}{c}(1 \mp \eta \cos\theta)$, in which $\theta = \arcsin(\frac{\lambda_0}{2n\pi}\frac{d\phi}{dx})$ while – and + are for anomalous refraction and reflection, respectively.

## Results

Upon the newly introduced forces, this study presents a metaphotonic approach to control the in-plane propulsion and out-of-plane levitation of metajets. Metajets are vehicles constructed from specially designed metasurfaces capable of manipulating light to drive or propel metajets themselves in desired directions. The mechanism of these metajets leverages unbalanced forces from anomalous refraction and reflection at a specific angle. As depicted in **Figure 1**(a), a linearly polarized beam illuminates the metajet and refracts (refection not illustrated) at an angle ($\theta$). The refracted light, in turn, imparts reactive metaphotonic forces on the metajet. The metaphotonic forces acting on the metajet result from the metasurface's phase difference per unit cell. Each unit cell of the metajet consists of a silicon cylindrical pillar (500 nm height) on a silicon dioxide base (100 nm thick) with a constant pitch (d$x$ = 450 nm) in the X-Y plane. The pitch and height of the silicon pillars are optimized to achieve the highest refraction efficiency at 1 μm wavelength.

**Figure 1**(b) delineates the generation of metaphotonic forces through anomalous refraction (left) and reflection (right) in a metajet. The input wavevector ($k_{in}$) propagates from the base of the metajet along the positive z-axis, indicating a zero in-plane wavevector ($k_{in}^{\parallel}$). The incident wave predominantly refracts (left) at an angle ($\theta$) upon engaging the metasurface, generating an output wavevector ($k_{out}$). This refracted wavevector emerges as the resultant of $k_{in}^{\parallel}$ and an effective wavevector ($k_{\phi}$), arising from a spatially distributed phase gradient with $k_{\phi} = \frac{d\phi}{dx}$. Despite the metasurface's high refraction coefficient, the balanced wave undergoes reflection (right) by the metajet, assuming minimal absorption. For both anomalous refraction (left) and reflection (right), the in-plane wavevectors follow the modified Snell's law ($k_{out}^{\parallel} - k_{in}^{\parallel} = k_{\phi}$). Therefore, under normal incidence ($k_{in}^{\parallel} = 0$) with $k_{out}^{\parallel} = k_{out} \sin\theta$, we can determine the direction of $k_{out}$ through $k_{out} \sin\theta = k_{\phi}$ and its magnitude with $k_{out} = \frac{2\pi}{\lambda_0} n$. With the direction and magnitude of wavevectors determined, we can follow our previous analysis to lay out the vector diagram for the generated forces. $F_{out}$ (blue) is antiparallel to $k_{out}$ (red), $F_{in}$ (blue) is parallel to $k_{in}$ (red), and very importantly, $F_{total} = F_{out} + F_{in}$ can define the $F_{total}$ (green). In reality, for obtaining the magnitude of $F_{out}$, we need to take the refraction and reflection efficiency ($\eta$) into account. Nevertheless, this analysis shows that the metaphotonic force is fully controllable by the efficiency ($\eta$) and phase gradient ($\frac{d\phi}{dx}$), all achievable by engineering the structure of metasurfaces.

Increasing the silicon pillar diameter alters the refracted light's phase, introducing a phase change. **Figure 1**(c) illustrates how increasing the unit cell pillar radii in the unit cell amplifies the phase difference between incoming and transmitted waves. Supercell design involves configuring inter-element phase changes ($d\phi$) for a 0 to $2\pi$ shift. Red markers indicate an 8-pillar configuration producing $\pi/4$ phase changes between adjacent pillars. Patterning supercells with constant unit cell pitch ($dx$) produces a saw-tooth phase change pattern, as **Figure 1**(d) illustrates. Each saw-tooth phase pattern consists of a phase change for a complete 0 to $2\pi$ phase shift depicting continuous refracted light at an angle ($\theta$). Light refraction at $\theta$ resulting from supercell phase changes is evident in the electric field distribution shown in **Figure 1**(e), highlighting the metasurface's ability to control and redirect electromagnetic waves. The photon momentum transfer from these redirected waves offers a promising mechanism for the propulsion of metajets.

Experimental analysis employing two distinct fabrication methods validates the 5-pillar supercell metajet's diffraction efficiency and motion. Refraction efficiency measurement utilizes silicon pillars etched on a prefabricated 500 nm silicon-on-silicon dioxide substrate stack. For motion studies, metajets are detached from a substrate using a sacrificial chromium layer on a silicon dioxide substrate, as **Figure 2**(a) illustrates. A transmission experimental setup (detailed in supplementary information) measures refraction intensity and momentum space beam deflection. **Figure 2**(b) depicts the beam shift from the center towards the m = +1 diffraction order, contrasting with the central reference beam spot. Post-processing of beam spot intensities yields plots where the grey line represents the reference beam spot, while the blue line illustrates the metasurface-refracted beam's diffraction intensity. The 58% diffraction intensity observed proves suitable for the initial metajet design iteration. **Figure 2**(c) showcases the metajet designed using the efficiency-measured metasurface. Placed in a liquid cell (described in the Methods section), the metajet interacts with a linearly polarized light beam incident from the cell's bottom. Five panels, each 10 seconds apart, record and display the metajet's motion. Notably, the metajet moves in the opposite direction of the refracted light, as indicated by the green arrow in **Figure 2**(b). This observation aligns with theoretical predictions, which posit that the metajet's propagation direction driven by the metaphotonic force, $F_{total}$, opposes the $k_\phi$. This experimental result corroborates the theoretical framework, demonstrating the intricate relationship between light refraction and the resulting propulsion forces acting on the metajet.

As discussed, the refraction angle affects the refraction vector and corresponding metaphotonic forces acting on the metajet. Modifying the phase change ($d\phi$) per unit cell ($dx$) enables precise manipulation of the refraction angle. The phase gradient increases from ($\pi/4$)/pitch to ($2\pi/3$)/pitch when transitioning from an 8-pillar to a 3-pillar supercell at constant pitch, amplifying the refraction angle. Empirical observations and simulations verify the correlation between phase gradient and the number of pillars in a supercell. **Figure 3**(a) presents refraction intensities at first-order diffraction modes (m) for metajets with 3 to 8-pillar supercell configurations. Maximum refraction consistently occurs in the m = +1 diffraction order, increasing intensity as pillar count decreases. SEM images in **Figure 3**(b) illustrate various supercell configurations, color-matched to the bar plots in **Figure 3**(a) and featuring a 1-µm scale bar. These findings indicate that metajets with a small number of pillars can lead to higher diffraction efficiency. Increasing the number of supercell elements while preserving pitch ($dx$) decreases the ratio $d\phi/dx$. Reducing the number of pillars in a supercell necessitates a higher phase change ($d\phi$) per unit cell pitch ($dx$) to maintain the complete 0-2$\pi$ phase shift for incoming light refraction. **Figure 3**(c) shows the refraction angle with varying pillars per supercell, keeping the pitch ($dx$ = 450nm) constant. A clear correlation exists between supercell configuration and refraction angle, with the 3-pillar design achieving the highest refraction angle. The phase change ratio $d\phi/dx$ significantly affects refraction angle and efficiency, offering a method to influence metaphotonic force and accelerate metajet movement. As shown in **Figure 3**(d), experimental observation of the metajet speed for each supercell configuration confirms the correlation between phase gradient and the speed of the metajet.

Our work presents a comprehensive motion analysis of metajets propelled by metaphotonic forces. Metajet movement, depicted three-dimensionally in **Figure 4**(a), reveals horizontal propulsion and vertical levitation. The horizontal propulsion results from the in-plane force arising from the momentum change in the phase gradient, whereas the vertical motion results from the out-of-plane component of the metaphotnoic force. The metajet demonstrates constant linear progression along the x-axis, with negligible movement in the y-direction. The 3D plot's numbered points correlate with metajet images in **Figure 4**(b), illustrating the device's positions along its trajectory. **Figure 4**(c) presents a 2D plot illustrating the metajet's distance traveled over time in both the x and z directions. Metajet displacement measurements, taken at consistent 5-second intervals in three dimensions, are displayed in **Figures 4**(a-c). Initially, the metajet rests freely in the sample cell, focused through the objective lens, shown in **Figure 4**(b) panel one. Activation of the laser beam from the sample cell's bottom induces metajet propulsion in x and z directions, as **Figure 4**(b) panel two illustrates. Positions 2, 3, and 4 in **Figure 4**(a-c) illustrate the metajet's z-direction movement, revealing an initial exponential acceleration due to radiative force, then stabilization as the metajet reaches the vertical limit of the sample cell. It can be observed in **Figure 4**(b) panel 4 that the metajet is entirely out of focus, displaying the propulsion state of the metajet, compared with **Figure 4**(b) panel 1. The out-of-focus image depicts that the metajet travels linearly in the x direction while levitating in the sample cell. The metajet exhibits linear x-axis displacement stemming from the unbalanced metaphotonic force. Manipulating the phase gradient ($d\phi/dx$) enables further characterization of the metaphotonic forces acting on the metajet in the x and z directions.

Manipulation is accomplished by varying the number of pillars per supercell from 3 to 8, illustrated in **Figure 4**(d). The forces are normalized to the power of the laser input. Using COMSOL software, the Maxwell stress tensor around a supercell configuration was analyzed to calculate the total optical force acting on the device. The highest $d\phi/dx$ ratio, observed in the smallest element supercell, correlates with maximum diffraction efficiency and optimal m = +1 order light refraction, resulting in optimal x-directional metaphotonic force. This experimental observation is in line with our theoretical claim of the in-plane metaphotonic force equation given by $F_x = -\frac{\eta p \lambda_0}{2\pi c}\frac{d\phi}{dx}$. The negative sign indicates that the force acts in the opposite direction of the phase gradient. In contrast, the magnitude of the in-plane metaphotonic force increases proportionally with the phase gradient ($d\phi/dx$) and the refraction efficiency ($\eta$). The top panel in **Figure 4**(d) reveals that the out-of-plane force ($F_z$) increases as the number of pillars ($N$) in a supercell increases. An increased $N$ leads to a decreased $d\phi/dx$ or $\theta$ while keeping the total phase in a supercell constant at ~2π. According to our previous analysis, we know that $F_z = \frac{np}{c}(1 \mp \eta \cos\theta)$, where – and + represent anomalous refraction and reflection. For anomalous reflection, because both the reflection $\eta$ (or R in **Figure S6**) and $\cos\theta$ increase with an increased $N$, it is easy to understand that $F_z$ also increases. On the other hand, the scenario in anomalous refraction is more complicated because the refraction $\eta$ (or T in **Figure S6**) decreases competing with the increased $\cos\theta$. For the scenario that $F_z$ increases as $\theta$ decreases to happen in anomalous refraction, the effect of decreasing refraction $\eta$ must dominate, recognizing the – sign in the $F_z$ equation. Ultimately, we must simultaneously consider anomalous refraction and reflection because they occur in experiments. The above analysis shows that enhancing the phase gradient ratio amplifies the metaphotonic force along the x-axis, boosting horizontal propulsion while reducing the force in the z-direction and limiting levitation. These findings highlight the controllability by precisely tuning the phase gradient and structures to manipulate horizontal and vertical forces, offering a pathway to advanced metaphotonic manipulation with sophisticated three-dimensional motions.

## Discussion

We have introduced metaphotonic manipulations to control the in-plane propulsion and out-of-plane levitation of metajets through metaphotonic forces supported by first-principles analysis. The metaphotonic forces are controllable via the spatially distributed phase gradient designed by deliberately arranged nanopillars. This engineering capacity using structures distinguishes the metaphotonic force from the optical gradient, scattering, and opto-thermal forces achieved by manipulating electromagnetic fields[11-14]. These optical forces that already have remarkable achievements in biology, particle physics, and chemistry at the microscopic and sub-microscopic scales are either parallel or perpendicular to the light propagation direction[15-20], while the metaphotonic force can be at any angle designed by structural engineering. Numerous more intriguing optical forces exist, such as optical pulling forces through judiciously controlled multipole scatterings or dedicatedly constructed shapes and backgrounds[21,22], but still need to demonstrate fully controlled motions of objects. Nevertheless, like enabling metasurfaces to manipulate all aspects (e.g., polarization, intensity, phase, and frequency) of light at will, structural engineering can lead to many new ways (e.g., engineering the dispersion for broadband responses and designing structures for high efficiency) to metaphotonic manipulations.

Moreover, leveraging the first-principles analysis, we directly integrate the phase gradient into the formula of metaphotonic force as the key controlling element. The formulated integration that reveals the out-of-plane levitation of metajets sets metaphotonic force apart from the most recent demonstrations of in-plane motion through structural engineering. These fascinating investigations have markedly advanced the field of optical force by controlling structures to mainly induce anomalous refraction in dimerized and asymmetric nanoantenna arrays[23], rotating nano-bar metasurfaces with spin-torque interactions[24], and spatially arranged plasmonic nanostructures employing linearly and circularly polarized light[25,26]. Regardless of structures, our first-principles theory could help understand the optical forces induced by anomalous refraction and reflection. Meanwhile, our theory shows that the metaphotonic force augments with increased optical power while having no limitations on the overall size of structured objects. Therefore, they may extend optical manipulation from the predominant microscopic and sub-microscopic scales to large settings, such as interstellar light sails in space exploration[27-30].

## Methods

### Simulation

The simulations employed FDTD software, Tidy3d, to model the unit cell, focusing on accurately capturing the behavior of light interacting with the metasurface. The process involved simulating a parametric sweep of various silicon pillar radii to analyze how changes in the pillar dimensions affected the phase shift of the plane wave. Tidy3d further simulated the electric field distribution, diffraction order, and efficiency across different supercell configurations to identify optimal design parameters. To assess the optical forces exerted on the metajets, COMSOL calculated the Maxwell stress tensor, providing detailed insights into the distribution and magnitude of these forces under various incident wave conditions.

### Fabrication

The sample to evaluate the refraction efficiency of the metasurface, 500 nm of amorphous silicon, is deposited on a silicon dioxide substrate using PECVD. Electron beam lithography (EBL) was performed by spin coating ZEP520A (at 2000 rpm for 60 seconds), a positive photoresist, to create cylindrical pillar patterns. The developed patterns were etched using Reactive Ion Etching (ICP at 600W for 3 seconds, followed by 400W for 2 minutes. The metajets fabrication process employed a similar top-down approach and extracted the metajets from the substrate by wet etching the sacrificial layer. The metasurface fabrication began with the deposition of a 50 nm chromium layer on a silicon dioxide ($SiO_2$) substrate using electron beam evaporation. Chromium is a sacrificial layer that extracts the metajets from the substrate, followed by depositing 100 nm of $SiO_2$ using plasma-enhanced chemical vapor deposition (PECVD) at 50nm/min. Subsequently, the PECVD process added a 500 nm amorphous silicon layer. Next, EBL was performed by spin coating ZEP520A (at 2000 rpm for 60 seconds), a positive photoresist, to create cylindrical pillars. Developer ZED-N50 was used to transfer EBL patterns on silicon, followed by a 30-second isopropyl alcohol (IPA) rinse. Reactive Ion Etching (RIE) is used to etch the patterns to form cylindrical silicon pillars (RIE forward power at 200W, ICP forward power at 400W, $SF_6$ at 15sccm, $CHF_3$ at 75sccm). The sample was then cleaned in Remover PG at 85°C for 15 minutes and rinsed with IPA. Shaping the metajet requires a second EBL step utilizing ZEP520A (at 1000 rpm spin-coating). ZED-N50 then develops the patterns for 2 minutes and 25 seconds, with an IPA rinse finalizing the process. The metajet shape was transferred to the substrate stack using the RIE etching (RIE forward power at 200W, ICP forward power at 400W, SF6 at 15sccm, CHF3 at 75sccm). Lastly, chromium etchant removed the sacrificial layer, freeing the metajets from the substrate. A final rinse in deionized water completed the fabrication process.

## Optical Measurements

The experimental setup for characterizing the metasurface comprises measurements in the momentum space to observe the diffraction efficiencies. A Chameleon Compact OPO VIS femtosecond pulsed laser generates a 1,000 nm light, which passes through a linear polarizer and is incident on the metasurface using a (40x Nikon 0.75NA) objective lens. The transmitted light is collected using an infinitely corrected collection (10x Mitotoyo 0.26NA) lens. Projecting refracted light from the metasurface onto the Fourier plane characterizes the light and exposes the diffraction modes, and their respective refraction efficiencies imaged with a CCD camera using the Amscope software. The refraction modes and their efficiencies are analyzed post processing the images of refracted light.

A vertical optical setup, outlined in the Supplementary Information section, enables analysis of the metajets' linear motion. Although detached from the substrate, the metajets still lie on the substrate. The substrate is placed in a small container and sonicated for five seconds to displace all the metasurfaces from the substrate and transfer them to the DI solution. A pipet delivers a few drops of the metajet solution to a glass slide featuring a spacer with a 1" diameter and 120µm thickness and covered with another glass slide. A sample cell with two glass slides facilitates the metajet linear motion observation. A beam splitter passes a pulsed laser and the lamp light to the metajet sample cell. The light is incident on the sample cell from the bottom using an objective lens (10x Mitotoyo 0.26NA). A laser beam spot of fifty µm width focuses on the metasurface, while manual stage movement keeps the beam spot on the metajet to observe continuous motion. The light from the sample cell is collected using a 10x objective lens and recorded using a CCD camera. Video post-processing analyzes the motion of the metajets. The vertical movement is calibrated by detecting the focal distances of the sample cell's top and bottom and measuring the distance from its base via camera focus adjustments, while the horizontal displacement is determined by analyzing the metajet's traveled distance through video post-processing.

**Fig. 1: Spatially distributed phase gradient for emerging metaphotonic force**

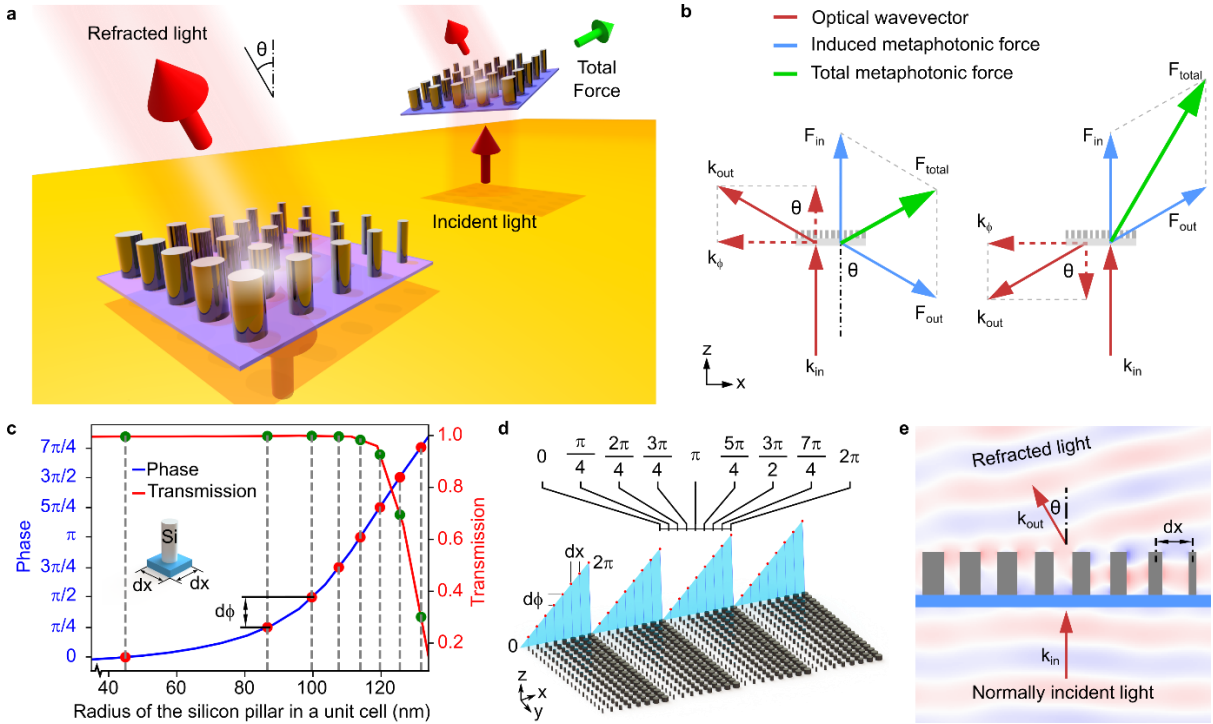

**a**, The schematic diagram of the metajet's motion (green) features a purposely arranged pillar configuration (not to the scale) that facilitates linear motion through refracted light (red) at an angle of θ (reflection not shown here). **b**, Optical momentum change represented by wavevectors gives rise to driving metaphotonic force. $F_{total} = F_{out} + F_{in}$, where $F_{total}$ denotes the total metaphotonic force, $F_{out}$ and $F_{in}$ are the force components generated by the output and input light. $k_{out}$ and $k_{in}$ are corresponding wavevectors, and $k_{out}$ is in different directions for anomalous refraction (left) and reflection (right). Governed by the modified Snell's law, the direction of $k_{out}$ represented by $\theta$ follows $k_{out} \sin\theta = k_\phi$, given $k_\phi = d\phi/dx$ under normal incidence light. **c**, Simulated transmission (red) and phase (blue) are functions of silicon pillar radius used to design $d\phi/dx$. Dots stand for the unit cells with a silicon pillar (height h = 500 nm) on a silicon dioxide base (thickness of 100 nm) for an 8-unit supercell in (**d**). The reduced transmission with increased radius is due to the higher filling factor of silicon in a unit cell. **d**, Phase gradient plot illustrating a gradual phase increase from 0 to 2π, forming a saw-tooth pattern for supercells along the x-direction. **e**, Simulated electric field distribution of refracted light on a supercell demonstrates light refraction at an angle of $\theta$. Note that the electric field distribution underneath is a superposition of the input and reflected light.

**Fig. 2: Maneuver of metajets with high optical diffraction efficiency**

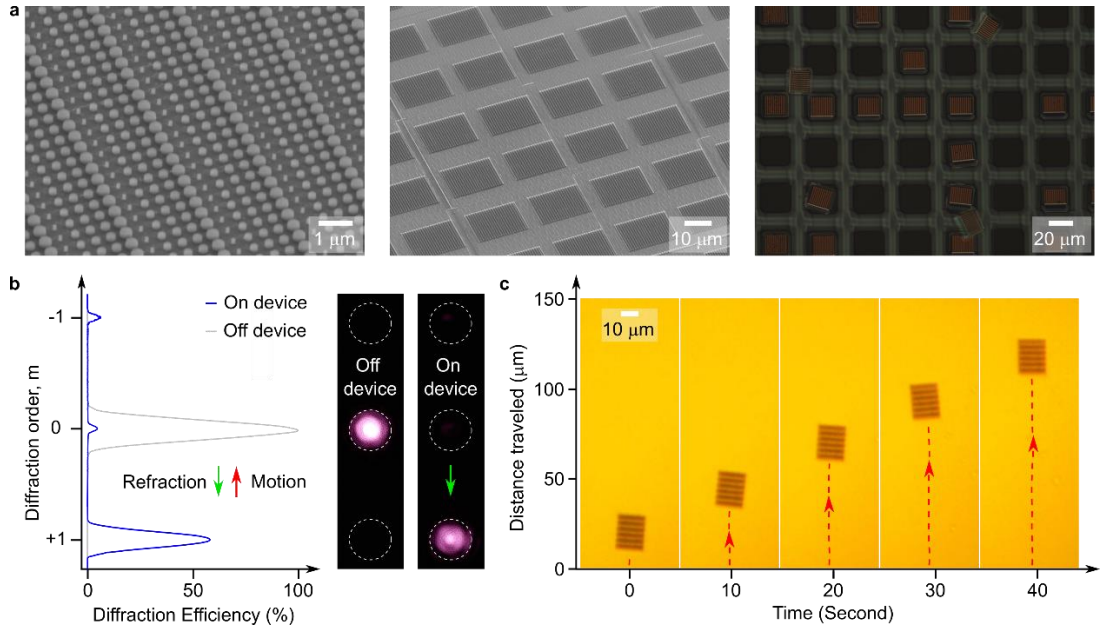

**a**, Scanning electron microscope (SEM) images elucidate the structure of silicon pillars (left), anchored metajets (middle) on a silicon dioxide ($SiO_2$) substrate, and free-standing metajets (right) after removing the sacrifice layer. **b**, Diffraction efficiency plots (left) compare refracted light (blue, on device) with an efficiency of 58% to a reference beam spot on the substrate (gray, off device). Momentum space imaging (right) reveals predominant light refraction in a diffraction order m = +1, corresponding to a diffraction direction (green arrow) opposite that of motion (red arrow) under the exact lab coordinates with (**c**). **c**, A time-lapse sequence demonstrates the linear motion of a metajet. The experiment employs a water-filled cell and a bottom-projecting 1-μm wavelength laser. The metajet's motion, opposite to the refracted light direction, results from reactive metaphotonic forces generated by the engineered metasurface with a phase gradient. Motion analysis reveals a speed of approximately 4.75 μm/s, benefiting from the high diffraction efficiency.

**Fig. 3: Manipulation of the metajets by characterizing phase gradients**

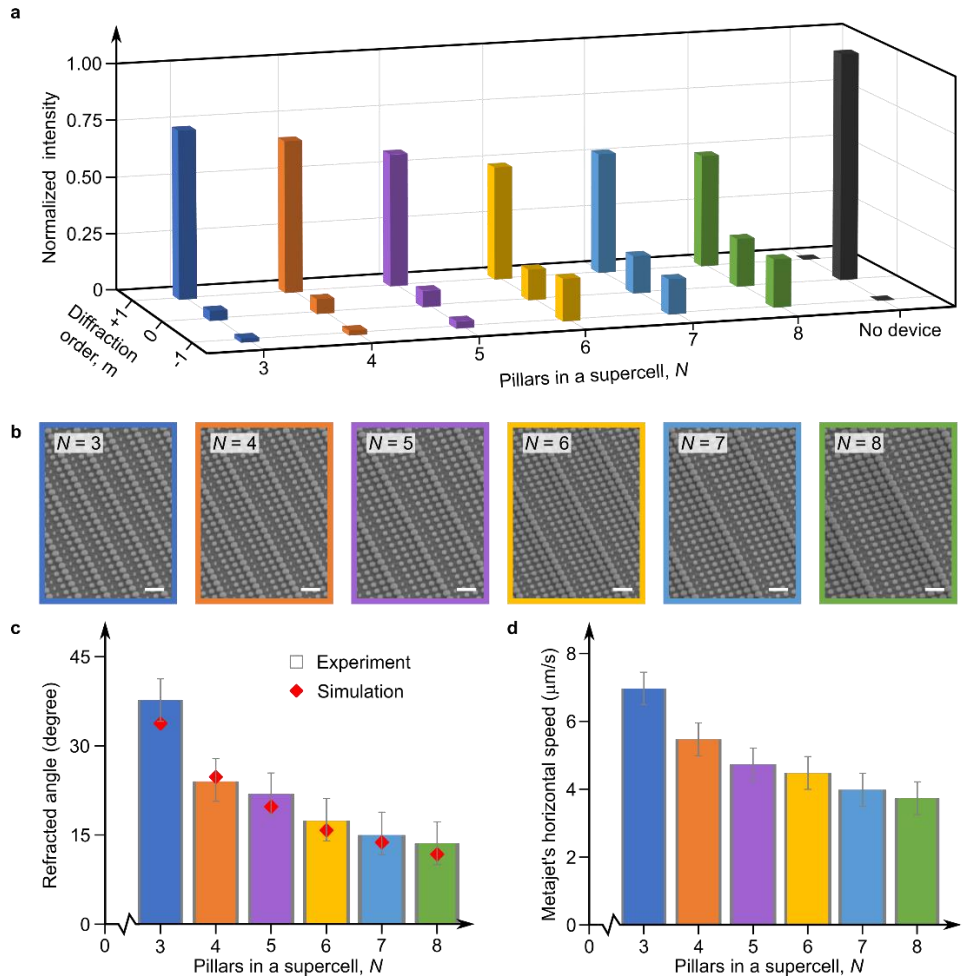

**a**, The experimental results reveal the metajet's controllable refraction efficiency by varying the supercell's pillar count (*N*) with a normalized intensity relative to a metasurface-free substrate. The 3-pillar supercell configuration exhibits the highest refraction efficiency (~78%) at the diffraction order m = +1. **b**, SEM images display the metasurface's cylindrical silicon pillars in configurations ranging from 3-pillar to 8-pillar supercells (from left to right), correlating with the experimental data in (**a**). The scale bar shown represents 1 µm. **c**, Experimental measurements of the variable refracted angle (corresponding to m = +1) match the trend observed in simulations (red diamonds). Decreasing the number of pillars in each supercell increases the ratio of $d\phi/dx$, as $dx$ (the pitch) remains constant while $d\phi$ (the phase change per pillar in the supercell) increases, assuming the total phase change in one supercell is roughly $2\pi$. **d**, Supercell configurations influence the metajet's speed, in which the 3-pillar supercell configuration exhibits the highest speed (~7 µm/s), resulting from its maximized refraction angle (~40° in **c**) and diffraction efficiency (~78% in **a**). Decreasing the number of pillars in a supercell leads to a lower speed, reflecting reduced refraction efficiency and refracted angles in the respective configurations. Data points are the average mean values, and error bars represent the standard deviation.

**Fig. 4: Demonstration of metaphotonic levitation**

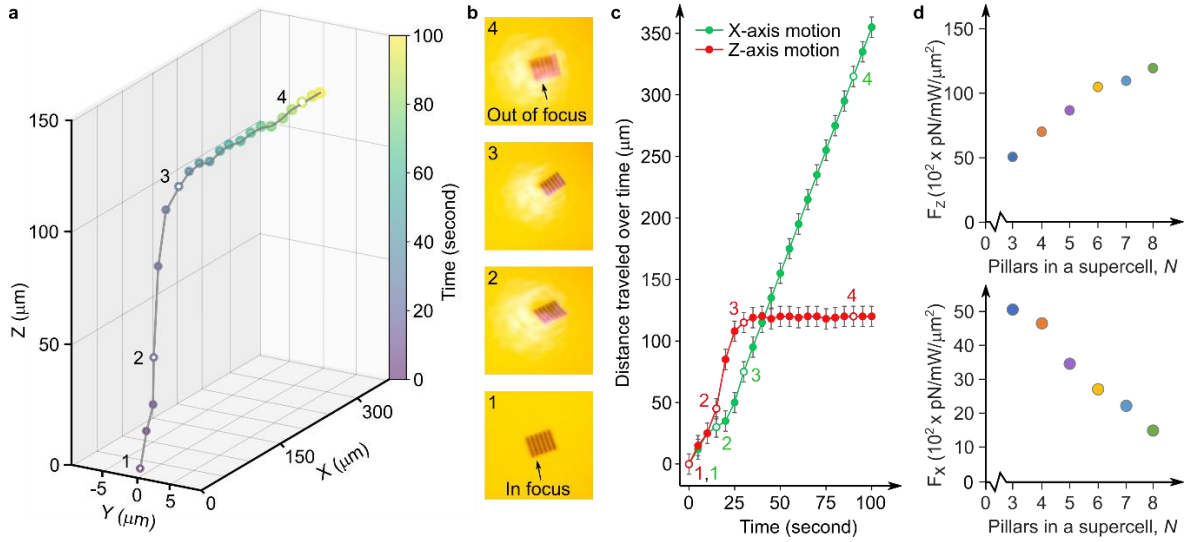

**a**, Experimental observation of the metajet's movement, demonstrating levitation induced by force $F_z$ and lateral motion caused by $F_x$. The initially focused state appears at point (1). Points (2) and (3) show the metajet's levitation in the z-direction and propulsion along the x-direction. Beyond point (3), the levitation distance reaches a constant level due to the container's cap used in the experiment, and the metajet progresses uniformly along the x-direction to point (4). **b**, Visual documentation illustrates the metajet's motion in (**a**). Levitation of the metajet causes a significant focus difference between (1) and (4), and the stage moves in the x-direction to keep the device in the field of view. **c**, Separate distance-time plots in x- and z-direction to better illustrate the metajet's initial 100-second movement. Data points represent the average mean values, and error bars are the standard deviation. **d**, Metaphotonic force simulations using the Maxwell Stress Tensor boundary. The normalized metaphotonic force for various supercell configurations highlights that the lateral force ($F_x$, bottom) peaks in 3-pillar configurations corresponding to the maximum refracted angle and refraction efficiency, while the vertical force ($F_z$, top) exhibits an inverse trend.